# Citation environment of *Angewandte Chemie*

Lutz Bornmann[*], Loet Leydesdorff, and Werner Marx

_______________________________________

[*]*Correspondence*: Dr. Lutz Bornmann

ETH Zurich

Professorship for Social Psychology and Research on Higher Education,

Zaehringerstr. 24

CH- 8092 Zurich

Tel.: +41 44 632 4825

Fax: +41 44 632 1283

E-mail: bornmann@gess.ethz.ch

**Abstract**

Recently, aggregated journal-journal citation networks were made accessible from the perspective of each journal included in the *Science Citation Index* (see http://www.leydesdorff.net/). The local matrices can be used to inspect the relevant citation environment of a journal using statistical analysis and visualization techniques from social network analysis. The inspection gives an answer to the question what the local impact of this and other journals in the environment is. In this study the citation environment of *Angewandte Chemie* was analysed. *Angewandte Chemie* is one of the prime chemistry journals in the world. Its environment was compared with that of the *Journal of the American Chemical Society*. The results of the environment analyses give a detailed insight into the field-embeddedness of *Angewandte Chemie.* The impacts of the German and international editions of this journal are compared.

**Key words**

# 1   Introduction

In 1976 Eugene Garfield – the founder of the Institute for Scientific Information (ISI, now Thomson Scientific, Philadelphia, PA, USA) – introduced the Journal Citation Reports (JCR) as an instrument to evaluate the significance of scholarly journals.[1] Today, the most important journals (currently about 7,500 journals from more than 3,300 publishers in over 60 countries) are listed in the JCR with a series of bibliometric data and indicators (e.g., total citations, Journal Impact Factor, Journal Immediacy Index, Journal Cited Half-Life). Through the publication of the indicators, essentially the Journal Impact Factor (JIF), JCR has become an authority for evaluating scholarly journals.[2, 3] To get from the JCR information on journals within disciplines for interdisciplinary comparison, each journal is classified by using 172 subject categories. For chemistry, the categories analytical, applied, inorganic/ nuclear, medicinal, multidisciplinary, organic, and physical are used (*Angewandte Chemie – International Edition*, *Angew Chem Int Edit*, belongs to multidisciplinary).

The JCR are an index of journal-journal links based on a grouping and summation of condensed citations using the journal name as the sorting key.[1] For each journal, there are two basic print-outs – cited and citing journals. Cited journal data show the number of times papers (original research and review articles) published in a certain year (in journals covered by JCR) cited papers (original research and review articles) published in a certain journal. Citing journal data show the number of times papers published in journals (covered by JCR) were cited in a certain journal in a certain year. Both data combined define a huge matrix of cited and citing journals.[4] From this huge matrix, local matrices can be extracted. Recently, local matrices were made accessible from the perspective of each journal covered by JCR (in 2004 from the perspective of 7,379 journals, see http://www.leydesdorff.net/jcr04). The local matrices can be used to inspect the relevant citation environment of one journal (i.e., the aggregated journal-journal citation network from the perspective of one journal) by using statistical analysis and visualization techniques from social network analysis.[5] The inspection may give answers to the following questions: what is the scientific impact of this and other journals in the environment (defined each by the number of published papers and their citations)? What is the share of the journal's self-citations among their local impact? Furthermore, and most importantly, environments can be divided in field-specific clusters by means of a journal's citation patterns; relationships between journals and clusters can be identified.

In this study the citation environment of *Angew Chem* (we refer to the German *and* international *Angewandte Chemie* editions with this abbreviation) was analyzed. *Angew Chem* is one of the prime chemistry journals in the world, with a JIF currently at 9.596 (JCR for 2005). The environment of *Angew Chem* was compared with that of the *Journal of the American Chemical Society* (*J Am Chem Soc*). *J Am Chem Soc* also belongs to multidisciplinary chemistry in the JCR with a JIF currently at 7.419 (JCR for 2005). Both *Angew Chem* and *J Am Chem Soc* are written and read primarily for the purpose of communication of original research findings.

## 2    Data and descriptive statistics

The data for the analysis of the *Angew Chem* environment originated from the 2004 journal-journal citation database of the JCR. In addition to the JCR data, we extracted the number of citations that *Angew Chem* received from other journals from the *Expanded Science Citation Index* at Web of Science (Thomson Scientific). *Angew Chem* has been published since 1962 in the form of both a German and an international edition, which has led to a number of problems with respect to citation analyses.[6, 7] Since *Angew Chem* papers are contained in different volumes and appear on different page numbers in the two editions, these papers are counted as if they were two completely different publications.[6] References to papers in *Angew Chem* relate to either one or the other edition or even both editions (double-citations), as practiced, e.g., for references in papers in *Angew Chem* to other *Angew Chem* papers.[6] As double-citations lead to an overestimation of the citations in JCR by the percentage of citations from both issues of the same paper,[8] we used *Angew Chem* data from the *Expanded Science Citation Index* that are corrected for double-citations in this study.

The local citation environment for *Angew Chem* was determined by including all journals which cited the journal to the extent of 1% of its citation rate: 76,904 (total citations to *Angew Chem Int Edit*) – 7,512 (self-citations to *Angew Chem Int Edit*) / 100 = 693 citations (on this, see the JCR for 2004). Among the 869 journals that cited *Angew Chem Int Edit* in 2004, only 22 cited it more than the 1% threshold of 693 times. Thus, the distribution of citations to the journal is highly skewed.[9] From the 22 journals that cited *Angew Chem Int Edit* more than 693 times all 852,008 references in 21,958 records were organized in a relational database. Among these 852,008 citations, 41,495 (5%) are to "Angew Chem" (both editions). One hundred and eleven of these records are incomplete (e.g., "In Press Angew Chem"); 32,416 (78%) of the remaining 41,495 records cited "Angew Chem Int Edit" and the other 9,079 (22%) "Angew Chem" without the indication of the international edition. We assume that this represents the German edition. In 8,054 cases, two consecutive records in the same document cited both the German and the international editions. When two citations follow upon each other, one to the German and the other to the international edition and with the same author name and publication year we identified these citations as double-citations. Among the 8,054 cases, we found 7,343 double-citations and 702 citations solely to the German edition. On the basis of this calculation, the impact of *Angew Chem* is overrepresented in this journal set (due to double-citations) to the extent of 21.5%: [7,343 / (41,495 – 7,343)] * 100. In a previous study[6], an overestimation of the (global) JIF with 15% was found, but the effect of 'double-citations' is larger in the *Angew Chem* citation environment because journals published by German publishers are overrepresented.

Table 1 shows the publisher and the subject categories of the 22 journals that determine the local citation environment of the *Angew Chem* as well as the total numbers of papers published in 2004 by each of the 22 journals. Seven journals are published by the American Chemical Society (Washington, DC, USA), five journals by Wiley - VCH Verlag (Weinheim, Germany), and three journals by the Royal Society of Chemistry (London, UK) and Pergamon (Amsterdam, The Netherlands), respectively. Most of the journals have been categorized by Thomson Scientific and Ulrich's International Periodical Directory[10] in the fields of multidisciplinary (five journals), inorganic/ nuclear (seven journals), and organic (12 journals) chemistry (many journals have been categorized in more than one field in both databases). Table 1 further shows that the number of papers published by the 22 journals averaged out about 1,015 with the highest number for *J Am Chem Soc* (3,167) and the lowest for *Chem Rev* (183).

Table 2 shows the number of papers published in 2004 by the 22 journals that cited papers published in *Angew Chem* (international edition, German edition, sum of both editions (double-citations included), and sum of both editions corrected for double-citations) in all years. The separated numbers (row 'Total citations') indicate that the international edition contributed about 70% (32,416) and the German edition about 30% (9,079) to the total number of citations.[6] However, there is a large amount of double-citations, especially produced by *Angew Chem Int Edit* (*n*=2,945) and *Chem-Eur J* (*n*=1,823). The table (column 'sum of both editions corrected for double-citations') shows that *Angew Chem* was cited 3,991 times (12%) by papers published in the same journal. That means about one tenth of citations from journals which contributed more than 1% to the *Angew Chem* impact are those of authors who published in *Angew Chem*. This within-journal citation rate of 12% is lower than, e.g., the rate of *J Am Chem Soc* (it has a share of about 20% within-journal citations among those journals that cited it more than 1% in the same year), but it is definitely higher than the within-journal citation rate of, e.g., *Chem Commun* (with much less than 10%). Although within-journal citations are often not self-citations by authors, they can be considered as an indicator of the inwardness of a community supporting a journal.[11] As Table 2 shows, 14% of citations to *Angew Chem* resulted from papers published in *J Am Chem Soc* and 7% from ones published in *Chem-Eur J* and *J Org Chem*, respectively. The remaining 60% of citations were provided by the other 18 journals listed in Table 2 (on average 1,138 citations per journal).

## 3 Results

For our set with journals that cited *Angew Chem* to the extent of 1% of its citation rate a 22 x 22 citation matrix was composed with two dimensions: one in the 'cited' (columns) and another in the 'citing' (rows). Both dimensions represent the relevant

*Angew Chem* citation environment (corrected for double-citations) covering the (local) impact of the most important journals. Each journal is represented in the matrix with the number of citations that it received from each of the other 22 journals ('cited' dimension) and with the number of citations that each of the other 22 journals received from this journal ('citing' dimension). The matrix was (1) imported into SPSS – a statistical analysis software[12] – for 'decomposition' of the environment with a reduction scheme, and (2) read into Pajek – a program for large network analysis[13] – for visualization of the environment (the visualization is based on using the algorithm of Tomihisa Kamada and Satoru Kawai).[14]

*'Decomposition' of the environment*

With exploratory factor analysis we tried to find a reduction scheme that indicates how the citation patterns ('being cited') of the 22 journals in the *Angew Chem* citation environment cluster or hang together. By the use of principal components in the factor analysis our set of correlated variables (the journals' citation patterns) was transformed into a set of uncorrelated variables (components or factors). The goal of this analysis was that a small number of components would account for most of the variance (>75%) in the patterns.[15] One result of the calculated factor analyses with four components (eigenvalues larger than 1) explained 80% of the variance and was taken as the final result (see Table 3).

The categorization of the journals among the components by using factor loadings greater than .4 (these loadings are marked in Table 3) showed that the journals can be categorized in terms of disciplinary affiliations (see also Table 1). A first group of journals (e.g., *Tetrahedron Lett* and *Synlett*) were cited in the environment according to a pattern of organic chemistry (factor 1 explains 40% of the variance in the matrix); a second and third pattern are specific for multidisciplinary chemistry (e.g., *J Am Chem Soc* and *Angew Chem*; also for applied chemistry: *Adv Synth Catal*) and inorganic/ nuclear chemistry (e.g., *Inorg Chem* and *Z Anorg Allg Chem*; factor 2 and factor 3 explain 25% and 9% of the variance, respectively). A fourth group can be designated to organometallic chemistry (e.g., *Z Organomet Chem* and *Organometallics*; factor 4 explains 6% of the variance).

As *Dalton T* and *Eur J Inorg Chem* embrace both aspects of the chemistry of inorganic and organometallic compounds, their citation patterns are not only specific for inorganic but also for organometallic chemistry. Because of low and negative factor loadings in Table 3, *J Phys Chem B* and *Org Biomol Chem* could not clearly be designated to one of the four factors. All in all, the results of the factor analysis show that the *Angew Chem* citation environment is determined by citation patterns that are characteristic for organic (factor 1), multidisciplinary (factor 2), and inorganic/ nuclear (organometallic, factor 3 and 4) chemistry journals.

*Visualization of the citation environment*

Figure 1 shows the citation environment of *Angew Chem* by using nodes for indicating a specific journal and solid lines for indicating links between journals. The links are defined by a similarity measure: the cosine coefficient.[16] This coefficient is similar to the well-known Pearson's correlation coefficient and specifies each correlation between the citation patterns ('being cited') of the 22 journals.[11] Links between journals representing cosine values below .2 are suppressed in Figure 1 in order to display only the most important correlations.

In agreement with the results of the factor analysis, the figure (in the centre) shows that *Angew Chem* is strongly embedded in a core group of five multidisciplinary chemistry journals (journals with red nodes). Organic and inorganic/ nuclear (together with organometallic) journals are connected to this core group at specific sites: organic chemistry journals at the top of Figure 1 (journals with green nodes) and inorganic/ nuclear chemistry at the bottom (journals with blue nodes). Also in agreement with the factor analysis findings, *J Phys Chem B* appears in the *Angew Chem* environment as a more or less isolated point. This journal presumably represents an environment according to a citation pattern of chemical physics and physical chemistry (surface chemistry, electrochemistry, and in particular the strongly increasing nanoscience). Almost 40% of the *J Phys Chem B* papers published in 2004 and citing *Angew Chem* (both editions) deal with topics related to nanoscience. At the bottom of Figure 1, *Z Anorg Allg Chem* clearly belongs to the inorganic journal group, but the node also appears somewhat isolated. We assume that *Z Anorg Allg Chem* is still noticed in the scientific community as one of the traditional German chemistry journals (and is not perceived as an international chemistry journal), although the number of papers published in English is currently larger than the number of papers published in German.

The shape of the nodes in Figure 1 is used for indicating the percentage contribution of each journal to the citation environment both including and excluding within-journal citations: the greater the vertical size of one node, the greater the share of received citations among the total number of citations in the environment. The share of received citations is not only dependent on the impact of the papers published by a certain journal but also by the total number of papers in the journal published in 2004;

both citations and number of articles strongly correlate (Spearman's rank correlation = 0.76). Whereas the vertical size of the nodes in Figure 1 indicates the share of citations including journal self-citations, self-citations are not considered for the horizontal size. The more ‚stretched' a node in the longitudinal axis appears, the higher is the share of self-citations. By inspecting the shape of the nodes in the figure one is able to see how much a journal is dependent on an inner-circle of authors citing one another. Note that within-journal citations can be both self-citations of authors and citations among authors publishing in the same journal.[11]

As the results in Figure 1 show, most of citations in the environment account for *J Am Chem Soc* with a share of 24% (19% without within-journal citations); about 8% account for *Angew Chem* (6% without within-journal citations). The high citation share for *J Am Chem Soc* compared to *Angew Chem* may especially be due to the fact that *J Am Chem Soc* published about 2.6 times more papers in 2004 than *Angew Chem* (see Table 1). Figure 1 further shows that *J Org Chem* and *Tetrahedron Lett*, repectively, received a share of about 10% (including within-journal citations). The shape of the nodes for these dominant journals in the environment indicates that the number of self-citations does not weight heavily: the citation patterns are not dominated by within-journal citation rates. The lowest number of citations in the *Angew Chem* citation environment account for *Org Biomol Chem* (0.2% with as well as without within-journal citations). If we sum the shares of the journals' citations according to the field categorizations in Table 3, the following percentage values arise: 47% for multidisciplinary chemistry (citation rate received by 7,411 papers published in seven journals), 38% for organic chemistry (citation rate received by 8,801 papers published in nine journals), and 12% for

inorganic/ nuclear and organometallic chemistry (citation rate received by 3,024 papers published in four journals).

For the purpose of comparison with the *Angew Chem* citation environment, the environment of *J Am Chem Soc* is shown in Figure 2. Among the 1,566 journals which cited *J Am Chem Soc* in 2004 at least one time, 21 cited it above the citation threshold of 1%. Whereas sixteen of the 21 journals appear also in the citation environment of the *Angew Chem* (see Figure 1), six journals drop out. Except for one journal (*Tetrahedron-Asymmetr*), these journals are published by German publishers (*Adv Synth Catal*, *Eur J Inorg Chem*, *Synlett, Synthesis-Stuttgart*, and *Z Anorg Allg Chem*; see Table 1). Five journals in the citation environment of *J Am Chem Soc* are not able to exceed the 1% threshold value in the *Angew Chem* environment: *Journal of Physical Chemistry A* (*J Phys Chem A*), *Journal of Chemical Physics* (*J Chem Phys*), *Macromolecules*, *Biochemistry* and *Langmuir*.

If we compare the visualizations for *Angew Chem* and *J Am Chem Soc* in Figure 1 and Figure 2 it is clearly visible that both citation environments are dominated by a strong core group of nearly the same journals with similar citation patterns. Journals which are each dropped out in the other citation environment are positioned in both figures primarily in the periphery around the core group. In the *Angew Chem* environment, the journals are to be found at different points in the periphery. However, in the *J Am Chem Soc* environment, these journals form a separate group beside the strong core group at the top in Figure 2. This means that these journals are characterized by similar citation patterns which are (more or less) different from those of the core group (except for *Biochemistry*). According to the subject categories provided by JCR and Ulrich's International Periodical Directory,[10] *J Chem Phys*, *J Phys Chem A*, and *J Phys Chem B* belong to the fields of physical chemistry and chemical physics. *Langmuir* is also strongly related to physical chemistry (e.g., surface and interface chemistry/ physics) and *Macromolecules* includes papers on surface properties of polymers.

## 4    Discussion

The JCR publish annual updated publication and citation data for measuring the performance of journals.[9] In this study the relevant citation environment of *Angew Chem* was analyzed by using these data (corrected for double-citations). The most important journals and their impact as well as field-specific journal-clusters were identified in the environment by means of statistical analysis and visualization techniques. The results of the environment analyses give a detailed insight into the field-embeddedness of *Angew Chem*. As data source JCR data for the year 2004 were analyzed. We assume that the results presented here for 2004 have time stability and can be extrapolated on 'flanking' years because a series of studies have shown remarkable time stabilities in journal citation patterns.[1, 17]

The 'decomposition' of a citation environment remains sensitive to the choice of the seed journal (the 'point of entrance'). Journal citation patterns span a multi-dimensional space in which clouds can be distinguished, but the delineation of these clouds at the edges remains fuzzy and varies with the perspective chosen by the analyst.[18] Therefore, the analyses of (inter-)disciplinary links between and among certain citation environments will be topics of great interest in future journal citation network analyses.

Table 1.

Journals that cited *Angew Chem* in 2004 more than 693 times: publisher, subject categories and total number of articles (published in 2004)

| Journal (abbreviated journal title) | Publisher | Subject category (from JCR and Ulrich's International Periodical Directory) | Number of articles |
|---|---|---|---|
| Advanced Synthesis & Catalysis (Adv Synth Catal) | Wiley - VCH Verlag, Germany | applied, organic chemistry | 223 |
| Angewandte Chemie-International Edition (Angew Chem Int Edit) | Wiley - VCH Verlag, Germany | multidisciplinary chemistry | 1,224 |
| Chemical Communications (Chem Commun) | Royal Society of Chemistry, UK | multidisciplinary chemistry | 1,321 |
| Chemical Reviews (Chem Rev) | American Chemical Society, USA | multidisciplinary chemistry | 183 |
| Chemistry-A European Journal (Chem-Eur J) | Wiley - VCH Verlag, Germany | multidisciplinary chemistry | 679 |
| Dalton Transactions (Dalton T) | Royal Society of Chemistry, UK | inorganic/ nuclear chemistry | 614 |
| European Journal of Inorganic Chemistry (Eur J Inorg Chem) | Wiley - VCH Verlag, Germany | inorganic/ nuclear chemistry | 577 |
| European Journal of Organic Chemistry (Eur J Org Chem) | Wiley - VCH Verlag, Germany | organic chemistry | 574 |
| Inorganic Chemistry (Inorg Chem) | American Chemical Society, USA | inorganic/ nuclear chemistry | 1,146 |
| Journal of The American Chemical Society (J Am Chem Soc) | American Chemical Society, USA | multidisciplinary chemistry | 3,167 |
| Journal of Organic Chemistry (J Org Chem) | American Chemical Society, USA | organic chemistry | 1,399 |
| Journal of Organometallic Chemistry (J Organomet Chem) | Elsevier, the Netherlands | organic, inorganic/ nuclear chemistry | 565 |
| Journal of Physical Chemistry B (J Phys Chem B) | American Chemical Society, USA | physical chemistry | 2,570 |
| Organic & Biomolecular Chemistry (Org Biomol Chem) | Royal Society of Chemistry, UK | organic, physical chemistry, biochemistry | 519 |
| Organic Letters (Org Lett) | American Chemical Society, USA | organic chemistry | 1,252 |
| Organometallics | American Chemical Society, USA | organic, inorganic/ nuclear chemistry | 875 |
| Synlett | Georg Thieme Verlag, Germany | organic, physical chemistry | 648 |
| Synthesis-Stuttgart | Georg Thieme Verlag, Germany | organic, physical chemistry | 472 |
| Tetrahedron | Pergamon, UK | organic chemistry | 1,203 |
| Tetrahedron Letters (Tetrahedron Lett) | Pergamon, UK | organic chemistry | 2,133 |
| Tetrahedron-Asymmetry (Tetrahedron-Asymmetr) | Pergamon, UK | organic, inorganic/ nuclear, physical chemistry | 555 |
| Zeitschrift fur Anorganische und Allgemeine Chemie (Z Anorg Allg Chem) | Wiley - VCH Verlag, Germany | inorganic/ nuclear chemistry | 426 |

Table 2.

Number of times papers published in 2004 (in journals below) cited papers published in *Angew Chem* (international edition, German edition, sum of both editions (double-citations included), and sum of both editions corrected for double-citations) in all years. The percent value specifies the share of a journal's citations among 'Total citations'.

| Journal | International edition | | German edition | | Sum of both editions (double-citations included) | | Sum of both editions corrected for double-citations | |
|---|---|---|---|---|---|---|---|---|
| | abs | % | abs | % | abs | % | abs | % |
| *J Am Chem Soc* | 4,757 | 15 | 264 | 3 | 5,021 | 12 | 4,846 | 14 |
| *Angew Chem Int Edit* | 3,485 | 11 | 3,451 | 38 | 6,936 | 17 | 3,991 | 12 |
| *Chem-Eur J* | 2,157 | 7 | 2,126 | 24 | 4,283 | 10 | 2,460 | 7 |
| *J Org Chem* | 2,315 | 7 | 203 | 2 | 2,518 | 6 | 2,378 | 7 |
| *Organometallics* | 1,866 | 6 | 276 | 3 | 2,142 | 5 | 1,942 | 6 |
| *Org Lett* | 1,903 | 6 | 113 | 1 | 2,016 | 5 | 1,938 | 6 |
| *Tetrahedron Lett* | 1,845 | 6 | 96 | 1 | 1,941 | 5 | 1,874 | 6 |
| *Inorg Chem* | 1,801 | 5 | 140 | 2 | 1,941 | 5 | 1,872 | 6 |
| *Tetrahedron* | 1,705 | 5 | 269 | 3 | 1,974 | 5 | 1,776 | 5 |
| *Chem Commun* | 1,681 | 5 | 109 | 1 | 1,790 | 4 | 1,721 | 5 |
| *Eur J Inorg Chem* | 1,148 | 4 | 371 | 4 | 1,519 | 4 | 1,195 | 3 |
| *Dalton T* | 1,047 | 3 | 138 | 2 | 1,185 | 3 | 1,092 | 3 |
| *Chem Rev* | 860 | 3 | 94 | 1 | 954 | 2 | 902 | 3 |
| *Eur J Org Chem* | 835 | 3 | 329 | 4 | 1,164 | 3 | 885 | 3 |
| *J Phys Chem B* | 811 | 2 | 66 | 1 | 877 | 2 | 868 | 3 |
| *Synlett* | 814 | 3 | 124 | 1 | 938 | 2 | 845 | 2 |
| *J Organomet Chem* | 610 | 2 | 143 | 2 | 753 | 2 | 680 | 2 |
| *Org Biomol Chem* | 689 | 2 | 42 | 0 | 731 | 2 | 692 | 2 |
| *Tetrahedron-Asymmetr* | 618 | 2 | 67 | 1 | 685 | 2 | 641 | 2 |
| *Synthesis-Stuttgart* | 543 | 2 | 118 | 1 | 661 | 1 | 569 | 1 |
| *Z Anorg Allg Chem* | 452 | 1 | 407 | 4 | 859 | 2 | 502 | 1 |
| *Adv Synth Catal* | 474 | 0 | 133 | 1 | 607 | 1 | 483 | 1 |
| Total citations | 32,416 | 100 | 9,079 | 100 | 41,495 | 100 | 34,152 | 100 |

Table 3.

Factor loadings resulting from a four factor solution (four factors explain 80% of the variance in the data)

| Journal | Factor 1: Organic chemistry | Factor 2: Multidisciplinary chemistry | Factor 3: Inorganic chemistry | Factor 4: Organometallic chemistry |
|---|---|---|---|---|
| *Tetrahedron Lett* | **0,940** | 0,134 | -0,194 | -0,125 |
| *Tetrahedron* | **0,933** | 0,148 | -0,206 | -0,132 |
| *Synthesis-Stuttgart* | **0,929** | 0,031 | -0,202 | -0,129 |
| *Synlett* | **0,920** | 0,061 | -0,242 | -0,135 |
| *J Org Chem* | **0,895** | 0,264 | -0,203 | -0,125 |
| *Eur J Org Chem* | **0,830** | 0,202 | -0,266 | -0,156 |
| *Org Lett* | **0,814** | 0,365 | -0,244 | -0,119 |
| *Tetrahedron-Asymmetr* | **0,530** | -0,123 | -0,427 | 0,028 |
| *J Am Chem Soc* | 0,081 | **0,948** | -0,062 | -0,004 |
| *Chem Rev* | 0,215 | **0,946** | -0,044 | 0,075 |
| *Angew Chem* | 0,168 | **0,915** | 0,019 | 0,080 |
| *Chem-Eur J* | 0,053 | **0,882** | 0,056 | 0,094 |
| *Chem Commun* | 0,189 | **0,867** | 0,241 | 0,225 |
| *Adv Synth Catal* | 0,346 | **0,429** | -0,565 | 0,313 |
| *Inorg Chem* | -0,284 | 0,327 | **0,772** | 0,211 |
| *Dalton T* | -0,346 | 0,197 | **0,714** | **0,461** |
| *Eur J Inorg Chem* | -0,345 | 0,219 | **0,664** | **0,545** |
| *Z Anorg Allg Chem* | -0,285 | -0,168 | **0,568** | -0,005 |
| *J Organomet Chem* | -0,173 | 0,170 | 0,073 | **0,882** |
| *Organometallics* | -0,210 | 0,288 | 0,031 | **0,863** |
| *J Phys Chem B* | -0,481 | 0,350 | -0,248 | -0,450 |
| *Org Biomol Chem* | 0,304 | 0,142 | -0,200 | -0,319 |

*Notes.* Extraction method: principal component analysis; rotation method: Varimax with Kaiser Normalization. Factor loadings greater than 0.4 are printed in bold, because they indicate a strong positive Pearson's correlation between the journal's citation pattern and the factor.[15]

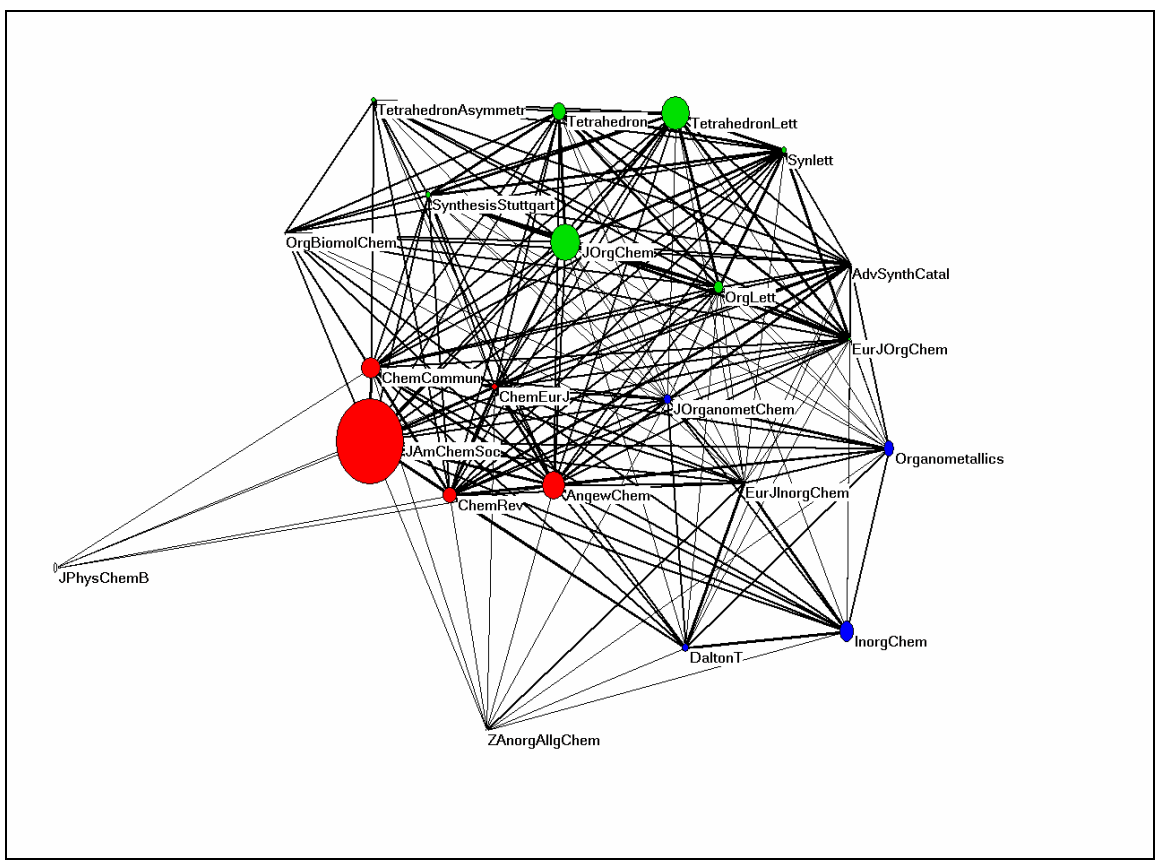

Figure 1.

Relevant citation environment of *Angew Chem* (corrected for double-citations). Share of journal's citations (without self-citations in parentheses) among total citations:

| | | | |
|---|---|---|---|
| *J Am Chem Soc* | 23.9% (19.0%) | *Synthesis-Stuttgart* | 1.9% (1.7%) |
| *J Org Chem* | 10.4% (8.5%) | *Synlett* | 1.9% (1.7%) |
| *Tetrahedron Lett* | 9.5% (7.9%) | *Tetrahedron-Asymmetr* | 1.7% (1.3%) |
| *Angew Chem* | 8.1% (6.3%) | *Eur J Org Chem* | 1.0% (0.9%) |
| *Chem Commun* | 6.0% (5.6%) | *Eur J Inorg Chem* | 0.8% (0.6%) |
| *Inorg Chem* | 5.8% (4.1%) | *Z Anorg Allg Chem* | 0.7% (0.4%) |
| *Tetrahedron* | 4.8% (4.1%) | *Adv Synth Catal* | 0.3% (0.3%) |
| *Organometallics* | 4.6% (3.0%) | *Org Biomol Chem* | 0.2% (0.2%) |
| *Chem Rev* | 4.3% (4.2%) | | |
| *Org Lett* | 3.6% (3.1%) | | |
| *J Organomet Chem* | 3.0% (2.4%) | | |
| *J Phys Chem B* | 2.8% (1.0%) | | |
| *Dalton T* | 2.7% (2.2%) | | |
| *Chem-Eur J* | 2.0% (1.8%) | | |

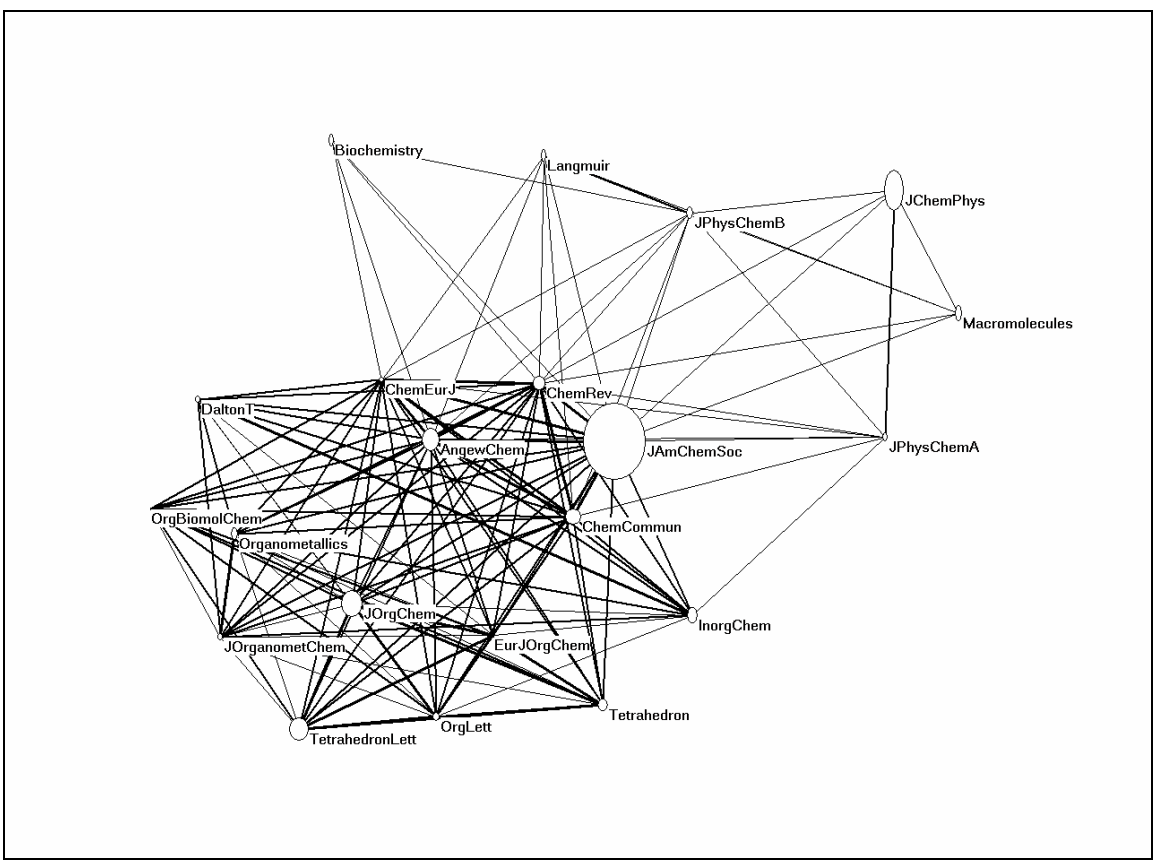

Figure 2.

Relevant citation environment of *J Am Chem Soc* (corrected for double-citations of *Angew Chem*). Share of journal's citations (without self-citations in parentheses) among total citations:

| | | | | |
|---|---|---|---|---|
| *J Am Chem Soc* | 21.7% (17.5%) | | *Tetrahedron* | 3.4% (2.8%) |
| *J Chem Phys* | 11.4% (5.5%) | | *Langmuir* | 3.3% (1.6%) |
| *J Org Chem* | 7.5% (5.9%) | | *J Phys Chem A* | 2.6% (1.6%) |
| *Tetrahedron Lett* | 6.7% (5.3%) | | *Org Lett* | 2.5% (2.1%) |
| *Angew Chem* | 6.4% (4.9%) | | *J Organomet Chem* | 2.2% (1.7%) |
| *Chem Commun* | 4.7% (4.3%) | | *Dalton T* | 1.9% (1.5%) |
| *Inorg Chem* | 4.4% (2.9%) | | *Chem-Eur J* | 1.6% (1.4%) |
| *Macromolecules* | 4.3% (1.8%) | | *Eur J Org Chem* | 0.7% (0.6%) |
| *Chem Rev* | 3.8% (3.7%) | | *Org Biomol Chem* | 0.2% (0.1%) |
| *Biochemistry* | 3.7% (1.5%) | | | |
| *Organometallics* | 3.5% (2.2%) | | | |
| *J Phys Chem B* | 3.5% (2.0%) | | | |